\documentclass[conference]{IEEEtran}
\IEEEoverridecommandlockouts
\usepackage{cite}
\usepackage{amsmath,amssymb,amsfonts}
\usepackage{algorithmic}
\usepackage{url}
\usepackage{graphicx}
\usepackage{textcomp}
\usepackage{xcolor}
\def\BibTeX{{\rm B\kern-.05em{\sc i\kern-.025em b}\kern-.08em
    T\kern-.1667em\lower.7ex\hbox{E}\kern-.125emX}}
\begin{document}

\title{Efficient training strategies for natural sounding speech
synthesis and speaker adaptation \\based on FastPitch
}

\author{
    \IEEEauthorblockN{Teodora RĂGMAN,\IEEEauthorrefmark{1}\IEEEauthorrefmark{2} Adriana STAN\IEEEauthorrefmark{1}\IEEEauthorrefmark{2}}
    \IEEEauthorblockA{\IEEEauthorrefmark{1}Communications Department, Technical University of Cluj-Napoca, Romania}
    \IEEEauthorblockA{\IEEEauthorrefmark{2}Speech and Dialogue Laboratory, POLITEHNICA Bucharest, Romania}
    \IEEEauthorblockA{\texttt{ragman.du.teodora@student.utcluj.ro, adriana.stan@com.utcluj.ro}}
}



\maketitle

\IEEEpubidadjcol

\begin{abstract}
This paper focuses on adapting the functionalities of the FastPitch model to the Romanian language; extending
the set of speakers from one to eighteen; synthesising speech using an anonymous identity; and 
replicating the identities of new, unseen
speakers. During this work, the effects of various configurations and training strategies were tested
and discussed, along with their advantages and weaknesses. Finally, we settled on a new configuration,
built on top of the FastPitch architecture, capable of producing natural speech synthesis, for both
known (identities from the training dataset) and unknown (identities learnt through short reference
samples) speakers. The anonymous speaker can be used for text-to-speech synthesis, if one wants
to cancel out the identity information while keeping the semantic content whole and clear. At last,
we discussed possible limitations of our work, which will form the basis for future investigations and
advancements.
\end{abstract}

\begin{IEEEkeywords}
speech synthesis, speaker adaptation, non-autoregressive architectures, speaker conditioning, anonymisation
\end{IEEEkeywords}

\section{Introduction}
The evolution of speech synthesis is beneficial in various fields with notable examples being virtual
assistants, medical use programs (for laryngectomized patients), and also language learning applications.
Modern text-to-speech (TTS) systems are neural network-based, continuing to push the boundaries set by their
predecessors. Most of the state-of-the-art TTS models are English based, and focus on replicating a single speaker’s
identity~\cite{matcha,lyth2024naturallanguageguidancehighfidelity}. If users are interested in replicating their own voice, they might decide on voice cloning models~\cite{wu23f_interspeech,sadekova22_interspeech}, where a
short reference (audio) sample is used to reproduce their voice based on a text prompt. Another important aspect of the speech
synthesis field addresses the privacy issues in speaker recognition and identity cloning applications, indicating the necessity of anonymous speakers. Usually, a new, sometimes robotic voice is used to mask and protect the user’s one while keeping the transmitted message intact. These three aforementioned tasks usually require the use of separate
models, each dedicated to its corresponding functionality~\cite{vali24_interspeech, meyer24_interspeech}. We propose extending the capabilities of a text-to-speech model to cover all three functionalities, starting with the FastPitch~\cite{Lancucki2021fastpitch} model as our baseline.

Our \textbf{main contributions} can be summarised as follows:
(i) altering the FastPitch model to allow multi-speaker, Romanian-based speech synthesis;
(ii) implementing the option of an ’anonymous’ speaker, which can be used in text-to-speech tasks to cancel out
speakers’ identities;
(iii) adapting the obtained TTS model to reproduce new, unseen speaker identities and use them for
speech synthesis tasks.

\section{Related Work}
\label{sec:rel}
Whereas early speech synthesis systems, such as WaveNet~\cite{oord2016wavenetgenerativemodelraw} and Deep Voice~\cite{arik2017deepvoicerealtimeneural} focused directly on generating raw waveforms, most of the latest TTS models prefer to split the synthesis process into two separate procedures. The first one aims to generate an intermediate acoustic representation, such as a Mel spectrograms, and the second one generates the time domain waveform. One of the most used acoustic generators is FastPitch~\cite{Lancucki2021fastpitch} known for parallel processing and handling long-range dependencies, which increases both the speed and generated content quality. Its architecture also served as inspiration for more advanced models, such as the zero-shot prosody cloning model from IMS-Toucan Toolkit~\cite{flux_prosodycloning}. IMS-Toucan leverages the FastPitch structure and adapts it to voice cloning tasks. Their work supports the importance of pitch, energy and duration information in obtaining expressive speech, along with the advantages of using Transformer models for natural language generation.

Even if these models have notable performances and generate accurate and natural-sounding samples, their functionalities are restricted to either text-to-speech synthesis or voice cloning. Also, models intended for privacy preservation are developed solely for this purpose, since irreversible identity masking can be a laborious process. This project explores the original structure of FastPitch in building a multi-speaker Romanian TTS speech synthesizer, with new functionalities, such as speech synthesis with an anonymous identity, and replicating unknown speakers’ voices based on short audio references. The last function was inspired by \cite{flux_prosodycloning} but requires minor alterations to the original structure and produces better samples. The introduction of the anonymous speaker component was driven by the importance of speaker identity information in predicting pitch, duration and energy parameters. By cancelling out the speaker embedding, we aim to study whether the generated samples exclusively lose speaker identity information, or if additional details, specific to the phrase and context, but not necessarily to the speaker, are also masked. While there are other options available for privacy protection~\cite{champion2024anonymizingspeechevaluatingdesigning} \cite{meyer2022speakeranonymizationphoneticintermediate}, we believe that this approach fully leverages the FastPitch structure, allowing a single base model to serve three distinct functions. This not only expands the capabilities of FastPitch, but also eliminates the need for separate models for these tasks.


\begin{figure}[t!]
    \centering
    \includegraphics[width=0.8\linewidth]{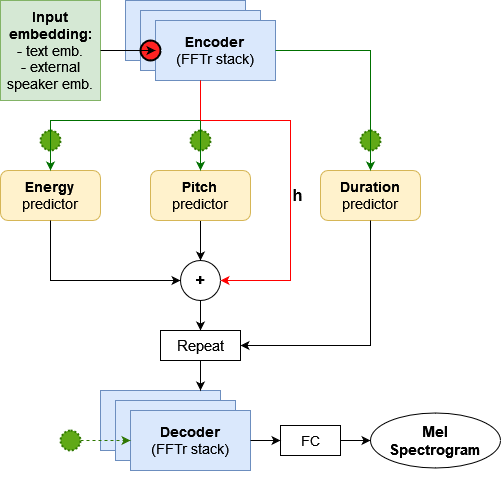}
    \caption{Simplified schematic of the FastPitch model. Circles represent conditioning points. The red circle indicates the  initial conditioning point of the original layout. There, the encoder is the only directly conditioned block.
    The green circles mark potential conditioning points. The three predictors (energy, pitch and duration) will be simultaneously conditioned. The decoder will be separately conditioned.}
    \label{fig:1}
    \vspace{-0.5cm}
\end{figure}

\section{Methodology}
\label{sec:meth}
Our main objectives consisted in adapting the FastPitch model for the Romanian language, expanding its speaker set, and incorporating features for speaker privacy preservation and speech synthesis with new identities. We believed that these three functions could effectively share one base architecture, and brought minimal alterations to improve the FastPitch performance, reduce computational efforts and shorten processing times.

We chose FastPitch because it allowed us to focus only on the acoustic generator and has been proven as a reliable structure for other more advanced models. Its structure is well divided into functional blocks, as illustrated in Figure~\ref{fig:1}, making it easy to bring minor alterations and study their effect. The most important components of the FastPitch structure are the two feed-forward Transformer blocks (\texttt{FFTr}), representing the encoder and the decoder. The encoder takes the character/phoneme embeddings as input and produces a hidden representation ($h$) used to predict the duration and pitch of each character. For every temporal location, a single pitch value will be predicted, and afterwards projected to match the dimensions of the hidden representation. These are then summed and up-sampled by repeated duplication according to the predicted duration. The result is fed to the decoder, which will generate the Mel spectrogram sequence. After this, any vocoder can be used to generate the synthesised waveform. During training, the model minimises the mean-squared error between predictions and ground-truth Mel spectrogram sequence, respectively pitch and duration.

\subsection{Multi-speaker Romanian TTS}
Our initial focus was on obtaining a stable TTS model, capable of generating Romanian speech with multiple speakers. FastPitch was initially designed for single-speaker, English-based speech synthesis, with the possibility to be trained on other languages and support multiple speakers. In its original layout, the model is designed to learn the speaker representations or embeddings, and reference this lookup table during inference. We modified this process to allow the initialisation of the speaker embedding layer from an external source, storing the embeddings of the speakers extracted using the TitaNet-L model\footnote{Available online: \url{https://catalog.ngc.nvidia.com/orgs/nvidia/teams/nemo/models/titanet_large}}. The speaker embedding layer is then frozen, allowing the model to use pre-computed representations during both inference and training. We trained the modified model on a Romanian language dataset~\cite{stan_sped2017}, obtaining a multi-speaker TTS model.

In the original FastPitch structure, the speaker embeddings are used to directly condition only the encoder block. The encoder output is then used as input for the other three predicting blocks, responsible for pitch, duration and energy, as shown in Figure~\ref{fig:1}. The green circles from the figure indicate where speaker embeddings can be used as a conditioning factor. Given that the three predictors (energy, pitch, and duration) and the decoder use the hidden representation produced by the encoder as input, they are indirectly conditioned on the speaker embedding. We wanted to test scenarios where the embeddings are not entered in the encoder, but instead directly and exclusively condition the other major blocks of the architecture. We trained three model configurations and varied the conditioning points, leading to conditioning: the \emph{encoder}, the \emph{predictors}, or the \emph{decoder} blocks. We then evaluated the last two against the first configuration–which follows the original structure.


\subsection{Speaker anonymisation}
The TTS task raised questions about the nature of the information carried by speaker embeddings -- whether they are related only to speakers’ identities or if the prosody, intonation and semantic elements affect the embedding values. To address this question, we considered a situation where the speaker information is removed during inference--would the text content alone be sufficient to capture the intended meaning of the phrase? To answer this question, we cancelled out the speaker information before using it as a conditioning factor and obtained an anonymous speaker identity. This voice serves as a common ground of all learned identities from the dataset and it is the only voice the model can generate in the absence of specific speaker information. We generated voice samples with this new identity using all three previously discussed configurations (obtained by varying the conditioning point). 

\subsection{Speaker adaptation}
Inspired by~\cite{flux_prosodycloning} where the FastPitch structure is modified to serve as a prosody cloning system, we pursued a speaker adaptation function to expand the output identities outside of the dataset speakers. Figure~\ref{fig:2} follows the steps a user must complete to adapt the TTS model to their own voice. The upper branch describes the embedding extraction process, where the denoised recordings are re-sampled and then processed by the TitaNet-L model. The lower branch focuses on generating custom metadata files – the re-sampled voice samples are input into an Automatic Speech Recognition (ASR)~\cite{radford2022robustspeechrecognitionlargescale} system, which provides transcriptions. These transcriptions are then converted into their phonetic representation using a pre-trained model and its associated lexicon~\cite{lorincz_nle_2022}. This phonetic version of the text is then used to create training and validation metadata files. The 22kHz voice samples are employed to generate the ground truth Mel spectrograms and pitch information, needed for further training steps.

The checkpoints used when resuming training are the three pre-trained TTS models--the encoder, predictors, respectively decoder-based configurations. The voice samples obtained with this new method (especially for the encoder and predictors conditioned configurations) sound natural and preserve the identity of the original speaker. However, finetuning the models adds computational complexity, and increases the risk of over-fitting, given that the training set is neither varied nor plentiful. One possible solution could be freezing the layers that are already well-optimised, method inspired by \cite{brock2017freezeoutacceleratetrainingprogressively}. This helps the model maintain the stability of the already useful and learned representations. 


\section{Experimental setup}
\label{sec:exp}

\subsection{Models}
The text-to-speech models required minimal modifications to the original structure. To integrate the Romanian language dataset we used the SWARA Speech Corpus~\cite{stan_sped2017}, which contains over 16 hours of parallel data from 18 speakers. The audio samples were re-sampled, grouped by speaker and passed to the TitaNet-L model for embedding extraction. The textual data was converted to phonemic transcription using a Romanian text processor and incorporated into the training and validation files. 

The training data consisted of re-sampled audio files along with the corresponding metadata files (containing the path to ground-truth files and the phonetic transcription of the spoken content). To explore three different configurations, we varied the conditioning points by adjusting the use of speaker information. Speaker embeddings were either passed as conditioning factors to the encoder or decoder, or simply added to the input of the three predicting blocks. In all cases, the effect is the same--the conditioning information is added to the input sequence before being processed by the model layers.

\begin{figure}
    \centering
    \includegraphics[width=0.8\linewidth]{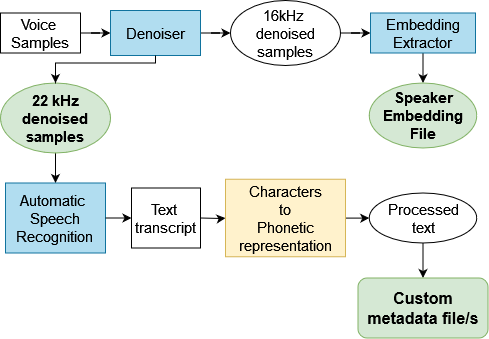}
    \caption{Simplified schematic of our pre-processing module. Green boxes mark the files making up the custom dataset. The upper branch describes the embedding extraction process, while the lower one focuses on generating custom metadata files.}
    \label{fig:2}
\end{figure}

To remove the speaker information and generate the anonymous identity, we multiplied the speaker embedding by 0 before using it as a conditioning factor. For speaker adaptation, the original dataset was replaced by a small custom dataset created for each new user. An interactive pre-processing module allows users to prepare their voice samples for the cloning process, its workflow being described in Figure~\ref{fig:2}. The training data now consists of the 22kHz (re-sampled) user recordings, along with the new custom dataset. Then, the training process is resumed for a few more steps, allowing the model to adapt to the new identity.

We evaluated our model’s capabilities using 30-second voice references. These recordings were re-sampled and used for embedding extraction. Training and inference were conducted on an NVIDIA T4 GPU, with a batch size of 16 for the base TTS models and 1 for fine-tuning the newly adapted models. The learning rate was set to 0.1, with the TTS models trained for 400k steps and model finetuning for 300 steps.

\subsection{Evaluation metrics}

The evaluation methods and metrics focused on the intelligibility of synthesised samples, along with our models’ ability to preserve the speaker identity. We experimented with directly conditioning the three predictors, respectively the decoder block with the speaker embedding and compared the results with the ones obtained for the encoder conditioned only configuration (the original setup of FastPitch).

Performance was measured using Word Error Rate (WER), Match Error Rate (MER) and Word Information Lost/Preserved (WIL/WIP) metrics from the Jiwer package,\footnote{\url{https://pypi.org/project/jiwer/}} based on ASR transcriptions of the generated and original samples. Speaker identity preservation was evaluated by computing the cosine similarity score of embeddings extracted from generated samples and their original references using the TitaNet-L architecture. The speaker adaptation task was evaluated using the same metrics, starting with the setup that did not involve layer freezing. Additionally, we considered the average Mel loss and the time required for additional training steps. We refer to these values when evaluating the selective layer freezing for the speaker adaptation.

\section{Results}
\label{sec:res}
We evaluated three functionalities – Romanian TTS, anonymous identity TTS and unseen speaker adaptation. For the Romanian TTS task, we used a test set of 10 new phrases that were not part of the training or validation sets. We generated voice samples from these new text prompts and evaluated the three Romanian text-to-speech configurations: encoder-based, predictors-based and decoder-based. Table~\ref{tab:1} summarises these results. It shows that the predictors conditioned only model achieved the lowest WER, nearly half the rate obtained for the original structure. Additionally, speaker identities were better preserved, as indicated by a cosine similarity score of $0.82$ when comparing the generated samples to their original references.

The experiments conducted for the anonymous speaker are based on the first four sentences from the test set used to evaluate the Romanian TTS function. We compared the voice samples generated with the anonymous identity to the original files carrying the same text. This functionality was added on top of the TTS structure and tested across all three conditioning configurations. Low objective measures should be indicating that this new identity is consistent across different phrases, produces clear and intelligible speech and bears minimal resemblance to the 18 identities from the training set. The lowest similarity score of 0.17 was obtained by the decoder-conditioned model, but this outcome was biased due to the less intelligible speech. This result was closely followed by the other two models, with scores of 0.18 for predictors and 0.2 for encoder based models, as listed in Table~\ref{tab:2}. Considering the trade-off, the predictors conditioned model yields a more reliable performance.

Focusing on the speaker adaptation function, we tested two fine-tuning approaches: one that simply resumes the training process, and another that involves selective layer freezing to enhance the performance. To evaluate these models, we used utterances from speakers unknown to our model (3 female and 3 male speakers). These 6 speakers are part of the SWARA2.0 extension of the training set, being recorded in rather similar conditions and uttering similar phrases. We also evaluated two novel speakers (one female and one male) unseen in neither SWARA, nor SWARA2.0 datasets. Table~\ref{tab:3} lists the results obtained for all 8 new speakers, using the first approach. The encoder conditioned model produced qualitative samples, but it was outperformed by the predictors conditioned model, which achieved a significantly lower WER. The decoder-based model performed poorly, and we decided to drop this configuration in the following tests.

\begin{table}[t!]
    \centering
    \caption{\textbf{Evaluation for seen speakers}. Average WER[\%], MER[\%], WIL[\%], WIP[\%] and cosine similarity for each TTS configuration. 10 unseen samples from each of the training set speakers were evaluated.}
    \begin{tabular}{lccccc}
         \textbf{Configuration} & \textbf{WER}$\downarrow$ & \textbf{MER}$\downarrow$ & \textbf{WIL}$\downarrow$ & \textbf{WIP}$\uparrow$ & \textbf{Cos Sim}$\uparrow$\\ \hline
         Decoder based & 21.09 & 19.77 & 28.87 & 71.13 & 0.56\\
         Encoder based & 20.74 & 19.23 & 28.87 & 71.13 & 0.62\\
         Predictors based & \textbf{11.28} & \textbf{10.26} & \textbf{15.10} & \textbf{84.90} & \textbf{0.82}\\ \hline
    \end{tabular}
    \vspace{-0.1cm}
   
    \label{tab:1}
\end{table}

\begin{table}[t!]
    \centering
    \caption{\textbf{Evaluation for the anonymous speaker}. Average WER[\%], MER[\%], WIL[\%], WIP[\%] and cosine similarity for 4 unseen phrases across the different TTS configurations and all SWARA speakers.}
    \begin{tabular}{lccccc}
         \textbf{Configuration} & \textbf{WER}$\downarrow$ & \textbf{MER}$\downarrow$ & \textbf{WIL}$\downarrow$ & \textbf{WIP}$\uparrow$ & \textbf{Cos Sim}$\downarrow$\\ \hline
         Decoder based & 13.42 & 12.61 & 17.64 & 82.35 & \textbf{0.17}\\
         Encoder based & 18.32 & 17.94 & 29.15 & 70.84 & 0.20\\
         Predictors based & \textbf{11.49} & \textbf{9.18} & \textbf{12.29} & \textbf{87.70} & 0.18\\ \hline
    \end{tabular}
    \vspace{-0.2cm}
    \label{tab:2}
    
\end{table}

\begin{table}[t!]
    \centering
    \caption{\textbf{Evaluation for unseen speakers (speaker adaptation)}. Average WER[\%], MER[\%], WIL[\%], WIP[\%] and cosine similarity for 8 unseen speakers across the different TTS configurations. 3 samples per speaker were used.}
    \begin{tabular}{lccccc}
         \textbf{Configuration} & \textbf{WER}$\downarrow$ & \textbf{MER}$\downarrow$ & \textbf{WIL}$\downarrow$ & \textbf{WIP}$\uparrow$ & \textbf{Cos Sim}$\uparrow$\\ \hline
         Decoder based & 80.19 & 69.57 & 82.07 & 17.93 & 0.63\\
         Encoder based & 38.00 & 31.46 & 43.68 & 56.32 & \textbf{0.68}\\
         Predictors based & \textbf{20.05} & \textbf{17.82} & \textbf{27.58} & \textbf{72.42} & 0.66\\ \hline
    \end{tabular}
    \vspace{-0.2cm}
    \label{tab:3}
    
\end{table}

\begin{figure}[ht!]
    \centering
    \includegraphics[width=0.65\linewidth]{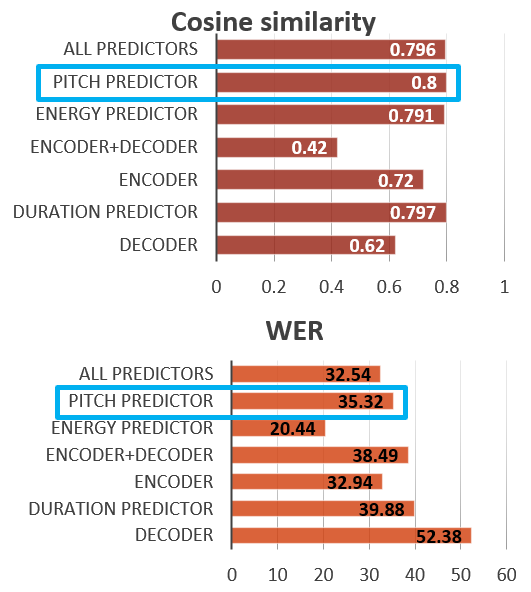}
    \caption{Cosine similarity ($\uparrow$) and WER [\%] ($\downarrow$) evaluation for the module freezing strategies. The results are based on the \textbf{encoder conditioned configuration}, using 30 second voice references from two unseen speakers.}
    \label{fig:3}
\end{figure}

\begin{figure}[ht!]
    \centering
    \includegraphics[width=0.65\linewidth]{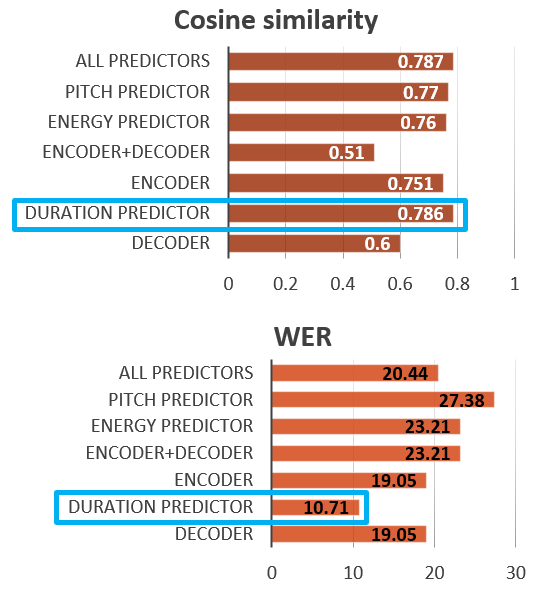}
    \caption{Cosine similarity ($\uparrow$) and WER [\%] ($\downarrow$) evaluation for the module freezing strategies. The results are based on the \textbf{predictors conditioned configuration}, using 30 second voice references from two unseen speakers.}
    \label{fig:4}
\end{figure}

 The second approach for the speaker adaptation process involved freezing one or multiple layers - the encoder, the decoder, all three predictors or one at a time, or both the encoder and decoder simultaneously. This yielded seven setups per configuration (encoder and predictors based). Figures~\ref{fig:3}~and~\ref{fig:4} highlight the best results and their corresponding setups. The encoder-conditioned model achieved the highest similarity score, but its WER values were quite high compared to the predictors-conditioned configuration. Considering how close the highest similarity scores obtained for each base structure are, and the large gap between their associated WER values, the predictors-based model fine-tuned with a frozen duration predictor offered the best performance. Moreover, the layer freezing strategy reduced the training time by nearly 33\%, while maintaining the average Mel loss values.

\vspace{-.2cm}
\section{Discussions}
\label{sec:disc}
The results obtained in the previous section show how different conditioning points affect the quality of synthesised speech, in terms of both identity preservation and discourse clarity. Samples obtained with the predictors conditioned only model proved to be the most qualitative, especially compared to the original FastPitch structure. By directly conditioning the three predicting blocks, we show that the speaker information weighs more in the prediction decisions, and helps the model obtain more realistic synthesised content. The same configuration also performed well in the anonymous speaker task.

Regarding the speaker adaptation task, both layouts offered interesting insights. The case where we resumed the training process for all layers yields satisfactory samples, but the WER values are high and the highest similarity score is lower compared to the values obtained with simple TTS synthesis. The alarming WER values along with the time required by the fine-tuning process encouraged us to improve this process through freezing certain layers. By applying this strategy, we allowed the other layers to extend their training on the new data whilst reducing the training time. WER values were drastically reduced from 20\% to approx. 11\%, and the corresponding similarity score of 0.78 improved upon the 0.68 score previously obtained, as shown in Figure~\ref{fig:4}. The new values are closer to the best results obtained while evaluating the text-to-speech module (Table~\ref{tab:1}), and the additional training time is also reduced by 33\%. 

Some of the largest limitations lie in the speaker privacy protection and adaptation fields, where the privacy preservation function does not consider the case where the anonymous speaker’s voice might resemble the voice of a potential user. This problem could be fixed by biasing the anonymous voice towards the speaker with the smallest similarity score to our user. Doing so would sacrifice the consistency of our anonymous speaker, but this method would ensure a better protection of privacy.

Another limitation regards the speaker adaptation function, where we tried to improve the quality of generated samples by using more utterances in the training process. We provided around 45 seconds of speech (instead of 25 to 30) and obtained superior quality, but we did not determine the minimum number of seconds needed for an accurate adaptation. Also, some imperfections in reference files were mirrored by the generated samples, such as clipped audio files or inconsistent speech duration.

\section{Conclusions}
\label{sec:conc}
In this paper we introduced our modified FastPitch structure for TTS and speaker adaptation tasks, and demonstrated that a single base model can serve three distinct functions. We adapted FastPitch to Romanian and added the option of using an anonymous identity for speech synthesis. This system leveraged our base structure's versatility to reproduce unknown speaker identities in synthesised speech. We experimented with various training strategies, finding a balance between coherence and identity preservation. The best results came from conditioning only the predicting blocks, responsible for the pitch, energy and duration information, rather than just the encoder as it was the case in the original structure. The selective layer freezing method also provided high quality results, reducing the training time and computational load while allowing the reproduction of new, unseen speaker identities. Considering the limitations caused by user provided samples and pre-processing steps, improvements can be made by optimising the speaker adaptation and anonymous identity functions--to be addressed in our future work.

\textbf{Acknowledgement.} This work was partially funded by EU Horizon project AI4TRUST (No. 101070190).

\bibliographystyle{IEEEtran}
\bibliography{mybib}
\end{document}